\documentclass[useAMS,usenatbib]{mn2e}
\usepackage{epsfig}
\usepackage{amsmath, amssymb,bm}

\title[The Warped Disc of NGC 4258]{The Warped Disc of NGC 4258}

\author[R. G. Martin]{Rebecca G. Martin \\ University of Cambridge,
  Institute of Astronomy, The Observatories, Madingley Road, Cambridge
  CB3 0HA\\}
\begin{document}

\date{}

\pagerange{\pageref{firstpage}--\pageref{lastpage}} 
\pubyear{2007}
\maketitle

\label{firstpage}

\begin{abstract}
  We consider the properties of the warped accretion disc in NGC 4258
  which is delineated by maser emission.  We use our analytical models
  to consider whether the disc could be warped by Lense-Thirring
  precession. We show that such models fit the shape of the disc well
  and we determine the goodness of fit for various combinations of the
  warp radius and the disc and black hole configurations. Though the
  fits are compelling evidence, we note that such a model has
  implications for the formation and longevity of the disc which might
  be problematic for the current understanding of Seyfert Galaxies.

\end{abstract}


\section{Introduction}

NGC 4258 is a bright barred-spiral galaxy at a distance of $7.2\,\rm
Mpc$ \citep{H99}. It is in Seyfert's first catalogue of active
galaxies \citep{S43}. It has anomalous spiral arms of $\rm H\alpha$
emission \citep{CC61} in the inner regions which are symmetric with
respect to the nucleus. This emission is probably from shocks formed
where the matter ejected from the nucleus meets the interstellar
medium \citep{K72, F86}.

Masers are the microwave equivalent of lasers. Water vapour maser
emission at a wavelength of $1.35\,\rm cm$ provides information about
the accretion discs around highly compact objects in the centres of
active galaxies. Masers are bright where there is no gradient in the
bulk line-of-sight velocity \citep{GG83,WW00}. In a nearly edge-on
Keplerian accretion disc the strongest features occur along the line
to the disc centre where the disc is moving perpendicular to the line
of sight, these are the systemic masers. We also see masers where the
plane of the sky intersects the disc, these are the high-velocity
masers.

Masers were first detected in the galaxy NGC 4258 by \cite{CHL84} and
\cite{H84}. The structure and dynamics of these discs can be
accurately probed with very long baseline interferometry because the
emission lines are strong and intrinsically narrow.  NGC 4258 has a
set of masers that might trace out a warped accretion disc which is
seen nearly edge on \citep{M95,H05}.  Radio interferometry has shown
that their rotation is Keplerian \citep{NIM93} and that the disc
extends from $2.8\,\rm mas$ to $8.2\,\rm mas$ from the nucleus.

At its distance of $7.2\,\rm Mpc$, in the plane of the sky we have the
scaling $1\,\rm mas=1.077\times 10^{17}\,\rm cm=0.0349\,\rm pc$.  The
Keplerian rotation of the high velocity masers implies an enclosed
mass of $3.78 \times 10^7\,\rm M_\odot$ within $2.8\,\rm mas$ of the
nucleus \citep{HGM96}. This is consistent with a central supermassive
black hole.

On a scale of milliarcseconds the central core breaks up into a jet
oriented along the axis of the water maser disc \citep{H97,H98}.  The
northern jet has a flux of $3\,\rm mJy$ and its mean location is about
$0.4\,\rm mas$ north (in the plane of the sky) of the implied position
of the black hole which is the centre of the rotation of the masers.
However the jet position varies in time.  The southern jet has not
varied in flux or position and is located about $1.0\,\rm mas$ south
of the black hole position. It has a flux of $0.5\,\rm mJy$. The
free-free absorption in the masing disc causes the difference in
brightness between the jets \citep{H97}. The position angles of the
jets are poorly determined. We estimate the northern jet is at a
position angle of $5^\circ \pm 15^\circ$ measured from \cite{H98}. The
jets do not appear to be very well aligned with each other and the
southern jet is at a position angle of about $173^\circ$. On larger
scales the jet appears to be orientated north-south \citep{C00}.

The systemic masers are about $0.57\,\rm mas$ below the disc centre as
indicated by the high velocity masers \citep{H05} and at a disc radius
of $3.9\,\rm mas$. Hence, the masers at the inner edge of the disc
suggest that it is inclined at an angle of $\zeta_{\rm
  in}=\sin^{-1}(0.57/3.9)\approx 8.4^\circ$ to the plane of the sky.
There appears to be no structure to the vertical positions of the
masers in the disc and so the disc is probably thin \citep{M95}.

Masers operate when the temperature of the surface layers in the disc
is greater than $300\,\rm K$ but less than $1000\,\rm K$ \citep{Mo95}.
The outer edge of the disc is where the molecular to atomic transition
occurs, just outside the outer maser.  The reason for the apparent
inner edge of the maser disc is not yet known.

The high-velocity masers and are not colinear with the systemic masers
and have negligible accelerations \citep{G95}. The radial dependence
of declination with respect to the systemic velocity of NGC 4258 is
anti-symmetric in the the red and blue-shifted masers \citep{M95}. It
is suggested that the rotation axis of the disc varies with radius by
an angle of up to $0.2$ radians \citep{HGM96} and so the disc is
warped.  \cite{H05} found that the maser spots show a deviation from
Keplerian rotation of about $9\,\rm km\,s^{-1}$. They modelled this
with a warped Keplerian accretion disc with a radial gradient in its
inclination of $0.034\,\rm mas^{-1}$.

There are several suggested explanations for the origin of the warp in
the disc. \cite{C06} considered some mechanisms for the warping and
precession of galactic accretion discs. \cite{PTL98} showed that it
could be produced by a binary companion orbiting outside the maser
disc. Such a companion would need a mass comparable to that of the
disc but there is no observational evidence for it.  A second
suggestion is that radiation pressure from the central black hole
produces torques on a slightly warped disc and the warp grows
\citep{P96}.  However the masing disc is stable against this radiation
instability if $\alpha_1 \le 0.2$ \citep{C07}, where $\alpha_1$ is the
viscosity parameter \citep{SS73}.

Alternatively, in the absence of other torques, \cite{C07} concluded
that the warping in the disc of NGC 4258 is due to the
Bardeen-Petterson effect. If we have an accretion disc around a
misaligned spinning black hole, Lense-Thirring precession drives a
warp in the disc. The inner parts of the disc are aligned with the
black hole \citep{BP}.

\cite{C07} found that the warp radius in the disc is comparable to or
smaller than the radius of the inner masers.  They find the timescale
of alignment of the system to be a few billion years. They based their
work on the results of \cite{SF} who assumed that the surface density
is constant. We use the more realistic power law surface density and
viscosities described by \cite{MPT07}. We fit these analytical disc
models, warped by the Lense-Thirring effect to the shape of the
observed maser distribution.

\section{Steady State Disc Model}

In this Section we consider some properties of accretion discs. We
assume that we have a steady state disc where $\nu_1 \Sigma=\,\rm
const$ and that the surface density is a power law
\begin{equation}
\Sigma=\Sigma_0\left(\frac{R}{R_0}\right)^{-\beta}
\label{sig}
\end{equation}
\citep{SS73}. To be in steady state, the viscosity must obey
\begin{equation}
\nu_1=\nu_{10}\left(\frac{R}{R_0}\right)^{\beta},
\label{viscs}
\end{equation}
where $\beta$, $\Sigma_0$ and $\nu_{10}$ are constants and $R_0$ is
some fixed radius.

According to \cite{SS73} the thickness of the disc in the gas pressure
dominated region has
\begin{equation}
H\propto r^{\frac{21}{20}}(1-r^{-\frac{1}{2}})^{\frac{1}{5}}
\end{equation}
where $r=R/3R_{\rm g}$ and $R_{\rm g}=2GM_{\rm BH}/c^2$.  This is well
approximated by $H/R=\,$const for $R\gg R_{\rm g}$. For NGC 4258
$R_{\rm g}=1.05\times 10^{-3}\,\rm mas$ which is much smaller that the
inner edge of the disc at $2.8\,\rm mas$ and so we assume that
$H/R=\,\rm const$ in our work.

\subsection{Viscosities}

There are two viscosities: $\nu_1$ corresponds to the azimuthal shear
(the viscosity normally associated with accretion discs) and $\nu_2$
corresponds to the vertical shear in the disc which smoothes out the
twist. The second viscosity acts when the disc is non-planar.  We
assume that the second viscosity obeys the same power law as $\nu_1$
so
\begin{equation}
\nu_2=\nu_{20}\left(\frac{R}{R_0}\right)^{\beta}
\end{equation}
and $\nu_{20}$ is a constant.

We use the $\alpha$-prescription to describe the viscosities in the
disc so that
\begin{equation}
\nu_1=\alpha_1 c_s H = \alpha_1 H^2 \Omega
\end{equation}
where $H$ is the scale height in the disc and $\Omega$ is the angular
velocity and similarly
\begin{equation}
\nu_2=\alpha_2 c_s H = \alpha_2 H^2 \Omega,
\end{equation}
where 
\begin{equation}
\alpha_1=\alpha_{10}\left(\frac{R}{R_0}\right)^x~~~{\rm and}~~~\alpha_2=\alpha_{20}\left(\frac{R}{R_0}\right)^x,
\end{equation} 
with $\alpha_{10}$, $\alpha_{20}$ and $x$ constant (see below) and
$R_0$ is some fixed radius which we define in Section~\ref{sec:rwarp}.
We generally take $\alpha_{10}=0.2$ and $\alpha_{20}=2$ \citep{LP07}.
The angular velocity in the disc is Keplerian so that
\begin{equation}
\Omega =\left(\frac{GM_{\rm BH}}{R^3}\right)^{\frac{1}{2}}.
\end{equation}
Because $c_{\rm s}=H\Omega$ the azimuthal shear viscosity is
\begin{align}
\nu_1  & = \alpha_1 \left(\frac{H}{R}\right)^2 R^2 \Omega \cr
 & = \alpha_{10}(GM_{\rm BH} R_0)^{\frac{1}{2}}  \left(\frac{H}{R}\right)^2 \left(\frac{R}{R_0}\right)^{x+\frac{1}{2}}.
\label{visc1}
\end{align}
This can be written in the form of equation~(\ref{viscs}) with
$x=\beta-\frac{1}{2}$ which is a constant and so
\begin{equation}
\nu_{10}= \alpha_{10}(GM_{\rm BH} R_0)^{\frac{1}{2}}  \left(\frac{H}{R}\right)^2 .
\end{equation}
We can find a similar equation for the vertical shear viscosity,
$\nu_2$.

\subsection{Surface Density}

We let the surface density be the power law given by
equation~(\ref{sig}) but we need to find $\Sigma_0$.  We have the
equation describing the hydrostatic equilibrium in the disc
\begin{equation}
\frac{dP}{dz}=-\rho g_z,
\label{dpdz}
\end{equation}
where $P$ is the pressure and $g_z$ is the $z$ component of the
gravitational force, $\rho$ is the density and the equation of state
is
\begin{equation}
P=\rho c_s^2,
\label{p}
\end{equation}
in the case of a vertically isothermal model  with $c_s$ the, constant
in $z$, sound speed.  We find the $z$-component of the gravitational
force to be
\begin{equation}
g_z=\frac{GM_{\rm BH}z}{(R^2+z^2)^{\frac{3}{2}}}\approx \frac{GM_{\rm BH}z}{R^3}
\label{gz}
\end{equation}
for $z \ll R$.  We can integrate equation~(\ref{dpdz}) over $z$ using
equations~(\ref{p}) and~(\ref{gz}) to find the pressure
\begin{equation}
P=P_0 \exp \left[ -\frac{z^2}{2}\frac{GM_{\rm BH}}{c_s^2 R^3}\right]
\end{equation}
and density
\begin{equation}
\rho=\rho_0 \exp \left[ -\frac{z^2}{2}\frac{GM_{\rm BH}}{c_s^2 R^3}\right],
\end{equation}
where $P_0$ and $\rho_0$ are the pressure and density at the midplane
$z=0$.  We can now find the surface density
\begin{equation}
\Sigma=\int_{-\infty}^\infty \rho \, dz=\sqrt{2 \pi} H \rho_0
\label{sig2}
\end{equation}
because
\begin{equation}
H=c_s \left(\frac{R^3}{GM_{\rm BH}}\right)^{\frac{1}{2}}.
\end{equation}

For a steady state accretion disc we have
\begin{equation}
\dot M=3 \pi \nu_1 \Sigma,
\label{mdot}
\end{equation}
with $\nu_1$ the usual viscosity.  We eliminate $\Sigma$ from
equations~(\ref{sig2}) and~(\ref{mdot}) to find
\begin{equation}
\rho_0 = \frac{\dot M}{3 \pi \sqrt{2\pi} \nu_1 H}
= \frac{GM_{\rm BH}\dot M}{3\pi \sqrt{2\pi}\alpha_1}\frac{1}{c_s^3 R^3}
\end{equation}
\citep{NM95} and, using equation~(\ref{sig2}), we find
\begin{equation}
\Sigma=\frac{\dot M}{3\pi \alpha_1}\left(\frac{R}{H}\right)^2\frac{1}{(GM_{\rm BH}R)^{\frac{1}{2}}}.
\end{equation}
If $H/R=\,\rm const$ then this is equivalent to
\begin{equation}
\Sigma=\Sigma_0 \left( \frac{R}{R_0}\right)^{-(\frac{1}{2}+x)},
\label{sigma}
\end{equation}
where $R_0$ is some fixed radius and
\begin{equation}
\Sigma_0=\frac{\dot M}{3\pi \alpha_{10}}\left(\frac{R}{H}\right)^2\frac{1}{(GM_{\rm BH}R_0)^{\frac{1}{2}}}.
\end{equation}
We have $\nu_1 \Sigma=\,\rm const$ which is necessary for a steady
state disc.

\subsection{Inclination of the Disc}
\label{SS}

Following \cite{MPT07} we consider the disc to be made up of annuli of
width $dR$ and mass $2\pi \Sigma R dR$ at radius $R$ from the central
object of mass $M_{\rm BH}$ with surface density $\Sigma(R,t)$ at time $t$ and
with angular momentum $\bm{L}=(GM_{\rm BH}R)^{1/2}\Sigma \bm{l}=L\bm{l}$.  The
unit vector describing the direction of the angular momentum of a disc
annulus is given by $\bm{l}=(l_x,l_y,l_z)$ with $|\bm{l}|=1$.

We use equation (2.8) of \cite{P92} setting $\partial \bm{L}/ \partial
t = 0$ and adding a term to describe the Lense-Thirring precession
(the last one) to give
\begin{align}
0=&\frac{1}{R}\frac{\partial}{\partial R}\left[ \left( \frac{3R}{L} \frac{\partial}{\partial R}(\nu_1 L)
  -\frac{3}{2}\nu_1\right)\bm{L}+\frac{1}{2}\nu_2RL\frac{\partial \bm{l}}{\partial R}\right] \cr
 & + \frac{\bm{\omega_{\rm p}} \times \bm{L}}{R^3}.
\label{main}
\end{align}
The Lense-Thirring precession is given by
\begin{equation}
\bm{\omega_{\rm p}} =\frac{2G\bm{J}}{c^2}
\label{omegap}
\end{equation}
\citep{KP85}, where the angular momentum of the black hole
$\bm{J}=J\bm{j}$ with $\bm{j}=(j_x,j_y,j_z)$ and $|\bm{j}|=1$ can be
expressed in terms of the dimensionless spin parameter $a$ such that
\begin{equation}
J=acM_{\rm BH}\left(\frac{GM_{\rm BH}}{c^2}\right).
\label{angmom}
\end{equation}

We use the frame of the black hole with the black hole at the origin
and with its spin along the $z$ axis so that $\bm{J}=(0,0,J)$. In this
frame, \cite{MPT07} find the steady state solution of
equation~(\ref{main}) for $W=l_x+il_y$, with $i=\sqrt{-1}$, to be
\begin{align}
W(R)=&\frac{2W_\infty}{\Gamma \left(\frac{1}{2(1+\beta)}\right)}\frac{(-i)^{\frac{1}{4(1+\beta)}}}{ (1+\beta)^{\frac{1}{2(1+\beta)}}}\left(\frac{R_0}{R}\right)^{1/4} \cr
&\times K_{\frac{1}{2(1+\beta)}}\left(\frac{\sqrt{2}}{1+\beta}(1-i)\left(\frac{R}{R_0}\right)^{-\frac{1+\beta}{2}}\right),
\label{Wwarp}
\end{align}
where $W_\infty$ is the constant value of $W$ as $R\rightarrow
\infty$. It represents the direction of the angular momentum of the
outer disc. This solution has the outer disc angular momentum vector
in the $x$-$z$ plane.  Here $K_\nu$ is the modified Bessel function of
order $\nu$ and $\Gamma$ is the gamma function. In deriving this
solution we assume that we can neglect the non-linear term
$\bm{l}.\partial^2\bm{l}/\partial R^2=-|\partial \bm{l}/\partial
R|^2$. We need the warping to be gradual enough for this to be true.

We let the outer disc have the fixed angular momentum direction vector
$(l_{x \infty},0, l_{z \infty})$. As $R\rightarrow \infty$, $W
\rightarrow W_\infty=l_{x \infty}=\sin \eta$ where $\eta$ is the
inclination of the outer disc to the black hole.  The inclination of
the disc to the black hole at radius $R$ (the $z$-axis) is
\begin{equation}
\theta(R)=\cos^{-1}(l_z)=\cos^{-1}(\sqrt{1-|W|^2}).
\end{equation}
We let the angle $\phi$ be the usual azimuthal angle in spherical
polar coordinates with $0\le \phi < 2\pi$. It is the angle in the
$x$-$y$ plane from the projected direction of the outer disc angular
momentum which in our model, given by equation~(\ref{Wwarp}), is along
the $x$-axis. 

\subsection{Disc Section}

The high velocity masers trace the disc in NGC 4258 in the plane of
the sky so we need to find the cross section of our disc model in this
plane. Our solution is in coordinates ($x$,$y$,$z$) in the frame of
the black hole and the outer disc lies in the $x$-$z$ plane.  We
initially take this to be aligned with the coordinates ($x_{\rm
  bh}$,$y_{\rm bh}$,$z_{\rm bh}$) which are the coordinates that we
take the disc section in the $x_{\rm bh}$--$z_{\rm bh}$ plane.

We rotate the coordinates ($x$,$y$,$z$) about the $z_{\rm bh}$ axis by
an angle $-\phi_{\rm c}$. The shape of the disc in the $x_{\rm
  bh}$--$z_{\rm bh}$ plane is given by
\begin{equation}
z_{\rm bh}(R)=- \frac{R}{R_0} \tan \theta \cos (\phi_{\rm c}-\phi_{\rm max}),
\end{equation}
where $\phi_{\rm max}=\arg (W)$ and we have the negative sign so that
our model has the same orientation as the maser data and
\begin{equation}
x_{\rm bh}(R)= \frac{R}{R_0} (\sin^2 (\phi_{\rm c}-\phi_{\rm max})+\cos^2 \theta \cos^2 (\phi_{\rm c}-\phi_{\rm max}))^{\frac{1}{2}}
\end{equation}
in the positive $x_{\rm bh}$ direction.  We are assuming that the disc
is symmetric so the side of the disc with negative $x_{\rm bh}$ can be
found by rotating this shape by $180^\circ$ about the $y_{\rm bh}$
axis.

If $\phi_{\rm c}=0$ we see the largest warp possible for a given disc
because we are taking a disc section in the plane which contains both
the black hole and the outer disc angular momentum.  Conversely, if
$\phi_{\rm c}=\pi/2$ then we see the least warping in our disc
section.

We next consider the disc section when the black hole is not in the
plane of the sky but inclined at angle $i_{\rm inc}$ to it. We rotate
the ($x$, $y$, $z$) coordinates about the $x_{\rm bh}$ axis through an
angle $-i_{\rm inc}$. For small $i_{\rm inc}$ the shape of our disc
becomes
\begin{equation}
z_{\rm bh}(R)=- \frac{R}{R_0} \tan \theta \cos (\phi_{\rm c}-\phi_{\rm max}) \cos i_{\rm inc}
\end{equation}
and $x_{\rm bh}$ remains the same. If the plane of the sky contains
the black hole spin then $i_{\rm inc}=0$. We can use this small angle
approximation because if the angle $i_{\rm inc}$ were big then we
wouldn't see the masers because they are only seen when there is no
gradient in the bulk line of sight velocity. This only happens when
the disc is close to edge on.

\begin{figure}
\centering
\epsfxsize=8.4cm
\epsfbox{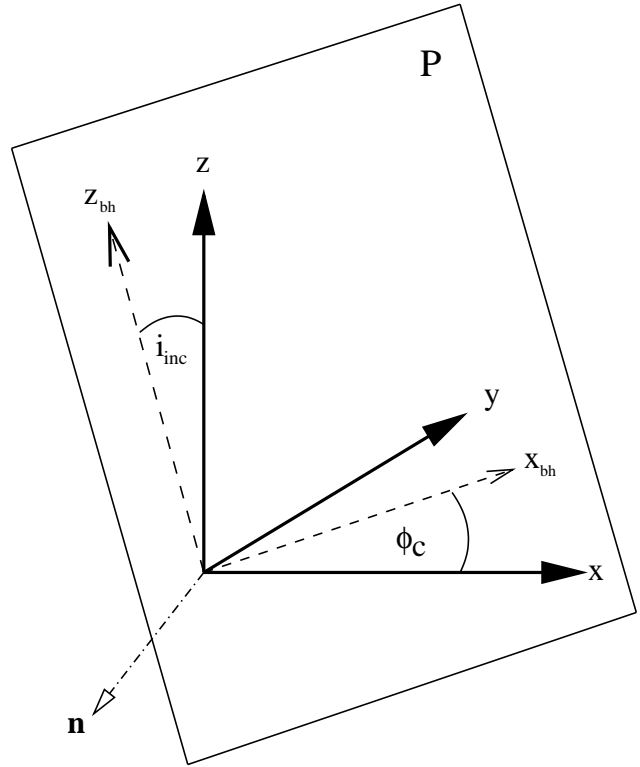}
\caption[]
{A diagram showing the plane of the sky, $P$, with normal $\bm{n}$
  which corresponds to the line of sight. We find the disc section in
  the plane $P$. The Cartesian coordinates used for the general
  solution to the warped disc equation given by equation~(\ref{Wwarp})
  are ($x$, $y$, $z$).  The black hole spin is up the $z$ axis and the
  outer disc angular momentum is in the $x$--$z$ plane.  The
  coordintates in the plane of the sky are $x_{\rm bh}$ and $z_{\rm
    bh}$, and $y_{\rm bh}$ is in the direction of $-\bm{n}$.  To find
  the orientation of the black hole and the disc we start with
  $(x,y,z)$ aligned with $(x_{\rm bh},y_{\rm bh},z_{\rm bh})$. We then
  rotate $x$ and $y$ about the $z_{\rm bh}$ axis by an angle
  $-\phi_{\rm c}$. Then we rotate about the $x_{\rm bh}$ axis by an
  angle $-i_{\rm inc}$.  The dashed lines are in the plane, $P$.  }
\label{fig}
\end{figure}

We sketch the plane of the sky coordinates ($x_{\rm bh}$, $z_{\rm
  bh}$) and the coordinate system for the disc ($x$, $y$, $z$) in
Figure~\ref{fig}. We show the plane of the sky, $P$, in which we find
the disc section.  It contains the dashed lines.  The normal,
$\bm{n}$, to the plane $P$ is the line of sight.  The disc section is
in coordinates $x_{\rm bh}$ and $z_{\rm bh}$.

In the case where $i_{\rm inc}=0$ and $\phi_{\rm c}=0$, we find the inclination
of the disc at radius $R$ to the plane of the sky to be
\begin{equation}
\zeta=\pi/2-\cos^{-1} (l_y).
\end{equation} 
The observations of the systemic masers predict that, at $R=3.9\,\rm
mas$, the disc has inclination $\zeta_{\rm in}=8.4^\circ$.  For our
models with $i_{\rm inc}=0$ and $\phi_{\rm c}=0$ we find the
inclination of our disc at $R=3.9\,\rm mas$ and compare this
to the observed value.

\section{Properties of NCG 4258}

The inner radius of the masing disc corresponds to $R_{\rm
  in}=2.8\,\rm mas$ and the outer edge to $R_{\rm out}=8.2\,\rm mas$.
\cite{A07} examine the systemic masers in the direction perpendicular
to the axis of the disc. They observe that the vertical thickness of
the disc has a limit $H<0.005\,\rm mas$.  Because the systemic masers
are at $R=3.9\,\rm mas$ we see there is an observed upper limit on
$H/R$ such that
\begin{equation}
\frac{H}{R}<0.0013.
\end{equation} 
We use a canonical value of $H/R=0.001$ in the following calculations.

\subsection{Warp Radius}
\label{sec:rwarp}

The warp radius is where the Bardeen-Petterson effect is balanced by
the viscous evolution of the disc and is given by
\begin{equation}
R_{\rm warp}= 2\frac{\omega_{\rm p}}{\nu_2(R_{\rm warp})}
\label{rw}
\end{equation}
\citep{SF,MPT07}, where $\omega_{\rm p}=|\bm{\omega}_{\rm p}|$.  Using
equation~(\ref{omegap}) and the form of equation~(\ref{visc1}) for the
second viscosity we find
\begin{equation}
R_{\rm warp}=\frac{4aG^2M_{\rm BH}^2}{c^3 \alpha_{20} R_{\rm warp}^2} \left(\frac{R}{H}\right)^2 
\left( \frac{R_{\rm warp}^3}{GM_{\rm BH}}\right)^{\frac{1}{2}}
\end{equation}
and, rearranging, 
\begin{equation}
R_{\rm warp}= \left[ \frac{4a(GM_{\rm BH})^{\frac{3}{2}}}{c^3\alpha_{20}}\left(\frac{R}{H}\right)^2\right]^{\frac{2}{3}}.
\end{equation}
More specifically, for NGC 4258 we find
\begin{equation}
R_{\rm warp}= 0.83 \, a^{\frac{2}{3}} \left(\frac{\alpha_{20}}{2}\right)^{-\frac{2}{3}}\left(\frac{H/R}{0.001}\right)^{-\frac{4}{3}}\,\rm mas.
\label{rwarp2}
\end{equation}
This warp radius is closer to the black hole than the inner masers and
we consider this in more detail in Section~\ref{sec:grav}.

\subsection{Accretion rate}

The motion of the infalling gas in NGC 4258 cannot be measured
directly so we have to estimate the mass accretion rate, $\dot M$,
indirectly. We use the bolometric luminosity of the system and assume
that
\begin{equation}
L_{\rm bol}=\epsilon \dot M c^2,
\end{equation}
where $\epsilon$ is the accretion efficiency of a Kerr black hole,
$\epsilon \approx 0.1$. The bolometric luminosity of NGC 4258 is about
$4\times 10^{42}\,\rm erg \, s^{-1}$. This corresponds to a mass
accretion rate of
\begin{equation}
\dot M= 7\times 10^{-5}\,\rm M_\odot \, yr^{-1}
\end{equation}
if the disc is in a steady state.

\subsection{Surface Density and Mass of the Disc}
\label{sec:mass}

We explore the consequences of the assumption that the disc is in
steady state.  When we vary $\beta$ we want properties of the disc at
$R=R_{\rm warp}$ to be be fixed and so we choose $R_0=R_{\rm warp}$.
We find the surface density using equation~(\ref{sigma}) to be
\begin{align}
\Sigma=\Sigma_0\left(\frac{R}{R_{\rm warp}}\right)^{-(x+\frac{1}{2})},
\end{align}
where
\begin{align}
\Sigma_0 & =  \frac{\dot M}{3\pi\alpha_1}\left(\frac{R}{H}\right)^2 \frac{1}{(GM_{\rm BH}R_{\rm warp})^{\frac{1}{2}}}\cr
& =  1.11 \times 10^2
\left(\frac{\alpha_{10}}{0.2}\right)^{-1}
\left(\frac{\alpha_{20}}{2}\right)^{\frac{1}{3}}
a^{-\frac{1}{3}}
\left(\frac{H/R}{0.001}\right)^{-\frac{4}{3}} \cr
&\times \left(\frac{\dot M}{7\times 10^{-5}\,\rm M_\odot \, yr^{-1}}\right)
 \,\ \rm g\, cm^{-2}.
\end{align}
\cite{C07} used the same model for the disc with $\alpha_1=\,\rm
const$ but varied $H/R$ with a power law in $R$. 

Using the surface density of the disc we can find the total mass of
the observable disc to be
\begin{equation}
M_{\rm d}=2\pi \int_{R_{\rm in}}^{R_{\rm out}}\Sigma R\,dR=\frac{2\pi \Sigma_0}{2-\beta} R_{\rm warp}^{\beta}(R_{\rm out}^{2-\beta}-R_{\rm in}^{2-\beta})
\end{equation}
if $\beta \ne 2$, or, if $\beta=2$ then
\begin{equation}
M_{\rm d}=2\pi \Sigma_0 R_{\rm warp}^{2}\log \left( \frac{R_{\rm out}}{R_{\rm in}}\right).
\end{equation}
For NGC 4258, with $\beta=2$, we find
\begin{align}
M_{\rm d} =  & 2.97 \times 10^3 \, a \left(\frac{\alpha_{10}}{0.2}\right)^{-1}
\left(\frac{\alpha_{20}}{2}\right)^{-1}
\left(\frac{H/R}{0.001}\right)^{-4} \cr
&\times \left(\frac{\dot M}{7\times 10^{-5}\,\rm M_\odot \, yr^{-1}}\right)\,\rm M_\odot.
\label{discmass}
\end{align}
This is much smaller than the mass of the black hole but in the next
section we consider if this is, or could be, large enough for our disc
to be self-gravitating.

\subsection{Self Gravity of the Disc}
\label{sec:grav}

Having calculated the mass of the observed maser disc we check if the
disc is stable against the effects of self gravity. If it were
unstable against self gravity then it would collapse into bound
fragments locally. If the mass of the disc satisfies the inequality
\begin{equation}
M_{\rm d} < M_{\rm BH} \left(\frac{H}{R}\right)
\end{equation}
then the disc is stable \citep{T64,BT87}. For $H/R=0.001$ this
corresponds to a critical disc mass of $3.78\times 10^4\,\rm M_\odot$.
For $\beta=2$, using equation~(\ref{discmass}), we see that we need
$H/R>0.0006$ for a disc which is stable against self gravity. Now we
have constrained $H/R$ to the approximate limits of
\begin{equation}
0.0006<H/R<0.0013
\end{equation}
if our model is correct. Using equation~(\ref{rwarp2}) we find that
for $\beta=2$ our warp radius is limited to
\begin{equation}
0.58<\frac{R_{\rm warp}}{1\,\rm mas}<1.65.
\label{warplimits}
\end{equation}
This is a relatively small range. We note that this range for $R_{\rm
  warp}$ is completely inside the inner observed maser points.
\cite{C07} found that the warp radius was approximately equal to or
smaller than the inner maser but we find that it is at most around
half way between the black hole and the inner maser.

We note that this mass just takes into account the mass of the disc of
the observed masers. For our steady state solution, the surface
density increases as we move towards the black hole and so, if we
calculate the mass of the disc including inner regions, then our disc
soon becomes self-gravitating. If the inner parts of the disc were not
in steady state or if $H/R$ were much larger there then we could over
come this problem.

\subsection{Gravitational Torques}

In our model, different radii of the disc communicate through the two
viscosities. Discs of non-negligible mass could also communicate
through gravitational torques.  We consider here how these torques
would affect equation~(\ref{main}). \cite{N05} considered the
gravitational torque between two rings at radii $R$ and $R_1$ of
masses $M$ and $M_1$ respectively which are inclined at an angle
$\gamma$ to each other. The torque exerted by the second ring on the
first is
\begin{equation}
T_{\rm grav}= \frac{GM_1 M}{4\pi^2} \int_0^{2\pi}  \int_0^{2\pi}
\frac{|\bm{r_1 \times r}|}{|\bm{r}-\bm{r_1}|^3}\,d\phi \, d\phi_1,
\end{equation}
where $\bm{r}$ and $\bm{r_1}$ are the position vectors around the
rings and the integration goes over $\phi$ and $\phi_1$ which are
azimuthal angles in the frame of the respective rings. He found the
maximum precession frequency to be when $R/R_1=\sqrt{3/7}$ and the
maximum torque to be
\begin{align}
T_{\rm grav} & = 0.085 \,  M \Omega^2 R^2  \sin \gamma  \cos \gamma \frac{M_1}{M_{\rm BH}}\cr
& = 0.085\, M \Omega^2 R^2  \sin \gamma \cos \gamma \frac{2\pi R_1 \Sigma_1 \delta R_1}{M_{\rm BH}},
\end{align}
where $M_1=2\pi \Sigma_1 R_1 \delta R_1$ is the mass of an inclined
annulus on which we consider the torque and $\Sigma_1=\Sigma (R_1)$.
In this formula $M$ is the mass of the annulus at $R$ which provides
the greatest torque. We replace $M$ by the mass of the disc, $M_{\rm
  d}$ to obtain an upper limit on the total torque on the annulus at
$R_1$.

The magnitude of the viscous torque is (equation~(\ref{main})
\begin{align}
T'_{\rm visc}& =\frac{1}{R_1^2}\frac{1}{2}\nu_2(R_1) R_1 L_1 \left|\frac{\partial \bm{l}}{\partial R}\right|_1 \cr
& = \frac{1}{2R_1^2}\nu_2(R_1) L_1 \sin \gamma
\end{align}
because
\begin{equation}
R_1 \left|\frac{\partial \bm{l}}{\partial R}\right|_1=\sin \gamma.
\end{equation}
We multiply by the area of the ring to find the viscous torque acting
on a ring of width $\delta R_1$ at radius $R_1$ to be
\begin{equation}
T_{\rm visc}=2\pi R_1 \delta R_1  \frac{1}{2R_1^2}\nu_2(R_1) L_1 \sin \gamma.
\end{equation}
We now compare the magnitude of the maximum gravitational torque to
the viscous torque
\begin{align}
  \xi_{\rm grav}= \frac{T_{\rm grav}}{T_{\rm visc}}& = \frac{0.085\,
    M_{\rm d} \Omega^2 R^2 R_1 \sin \gamma \cos \gamma \frac{2\pi
      R_1 \Sigma_1}{M_{\rm BH}}} {\pi \nu_2(R_1) L_1 \sin \gamma} 
\end{align}
and subsituting we find
\begin{align}
  \xi_{\rm grav}= \frac{T_{\rm grav}}{T_{\rm visc}}& = \frac{0.085\,
    M_{\rm d} \Omega^2 R^2 R_1 \sin \gamma \cos \gamma \frac{2\pi
      R_1 \Sigma_1}{M_{\rm BH}}} {\pi \nu_2(R_1) (GM_{\rm
      BH}R_1)^{\frac{1}{2}} \Sigma_1 \sin \gamma} \cr & =
  \frac{0.17\, M_{\rm d} G R_1^2 \cos \gamma} { \nu_2(R_1) R (GM_{\rm
      BH}R_1)^{\frac{1}{2}}} \cr & = \frac{0.17\, M_{\rm d} R_1^2 \cos
    \gamma} { \nu_{20} R (G^{-1}M_{\rm
      BH}R_1)^{\frac{1}{2}}}\left(\frac{R_1}{R_{\rm warp}}\right)^{-
    \beta} .
\end{align}

With $\beta=2$ and $\cos \gamma \approx 1$ at $R=R_{\rm in}$ and
$R_1=\sqrt{7/3}R_{\rm in}$ we find that $\xi_{\rm grav}= 3 \times
10^{-4}(M_{\rm d}/{\rm M_\odot})$ and at $R_1=R_{\rm out}$ and
$R=\sqrt{3/7}R_1$ then $\xi_{\rm grav}=10^{-4}(M_{\rm d}/{\rm
  M_\odot})$. Using the disc mass derived in Section~\ref{sec:mass} of
$M_{\rm d}=3 \times 10^3 \,\rm M_\odot$ we find at $R=R_{\rm in}$,
$\xi_{\rm grav}= 0.9$ and at $R_1=R_{\rm out}$, $\xi_{\rm grav}=0.3$.
This disc mass is the critical one at which the gravity torques become
as important as the viscous torques. For smaller disc masses self
gravity torques would be unimportant. In any case the gravitational
torque is not as large as the viscous torque even at its maximum so we
are justified in neglecting it.  However, it ought to be included in
any numerical models especially for larger disc masses.

\section{Comparing the Model to Observations}
\label{sec:comp}

In this section we compare the maser data to our analytical disc
models and find parameters which give the best fits.

\subsection{Maser Data}

We use the maser distribution data tabulated in the online version of
\cite{A07}. This lists positions of each maser in the plane of the
sky, $z_{\rm m}$ and $x_{\rm m}$. For the first set of data, BM056C,
we correct the north-south position of the masers (Alice Argon,
private communication). They correlated the first three epochs at a
less accurate position than later epochs and the data in the online
table for BM056C reflect this.  From the relativistic velocity,
$v_{\rm rel}$, in column~3 of Table~5 in the online table of
\cite{A07}, we find the frequency
\begin{equation}
f_{\rm freq} = 22.23508 \left( \frac{1 - v_{\rm rel}/c}{1 + v_{\rm rel}/c} \right)^{1/2}  {\,\rm GHz},
\end{equation}
where $c = 2.998 \times 10^{10}\,\rm cm\,s^{-1}$. Then we apply the
correction
\begin{equation}
 z_{\rm m} \rightarrow z_{\rm m} - 66.0 \left( \frac{f_{\rm freq}}{22.19856}-1 \right) \,\rm mas 
\end{equation}
and $x_{\rm m}$ remains the same.

\begin{figure}
\epsfxsize=8.4cm 
\epsfbox{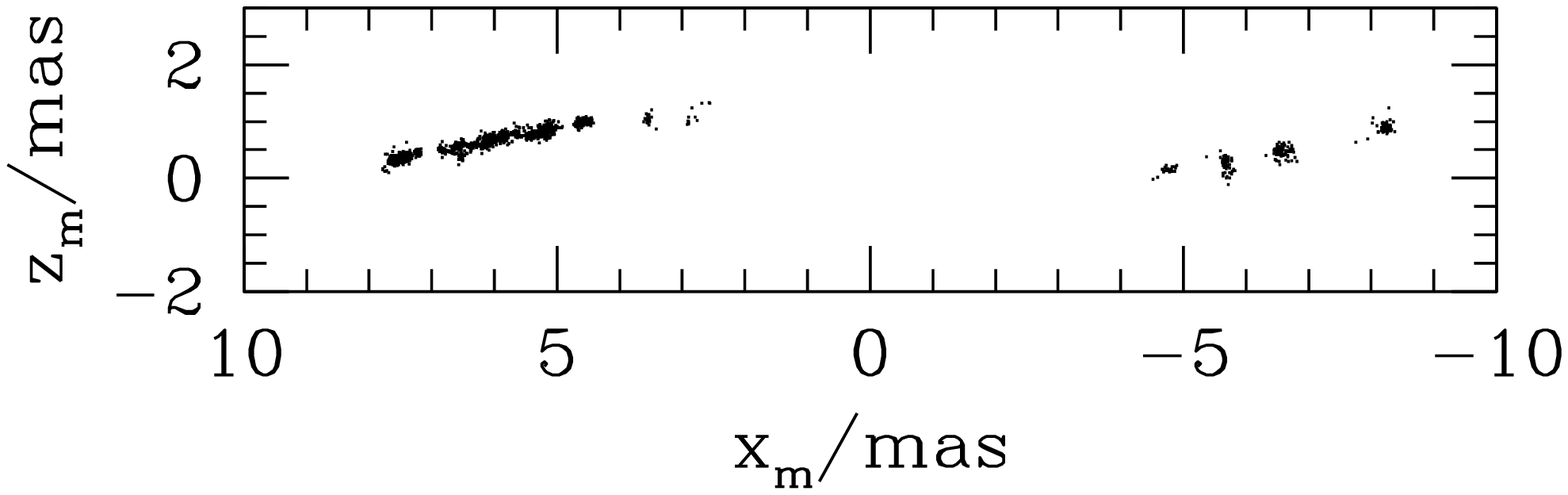}
\epsfxsize=8.4cm 
\epsfbox{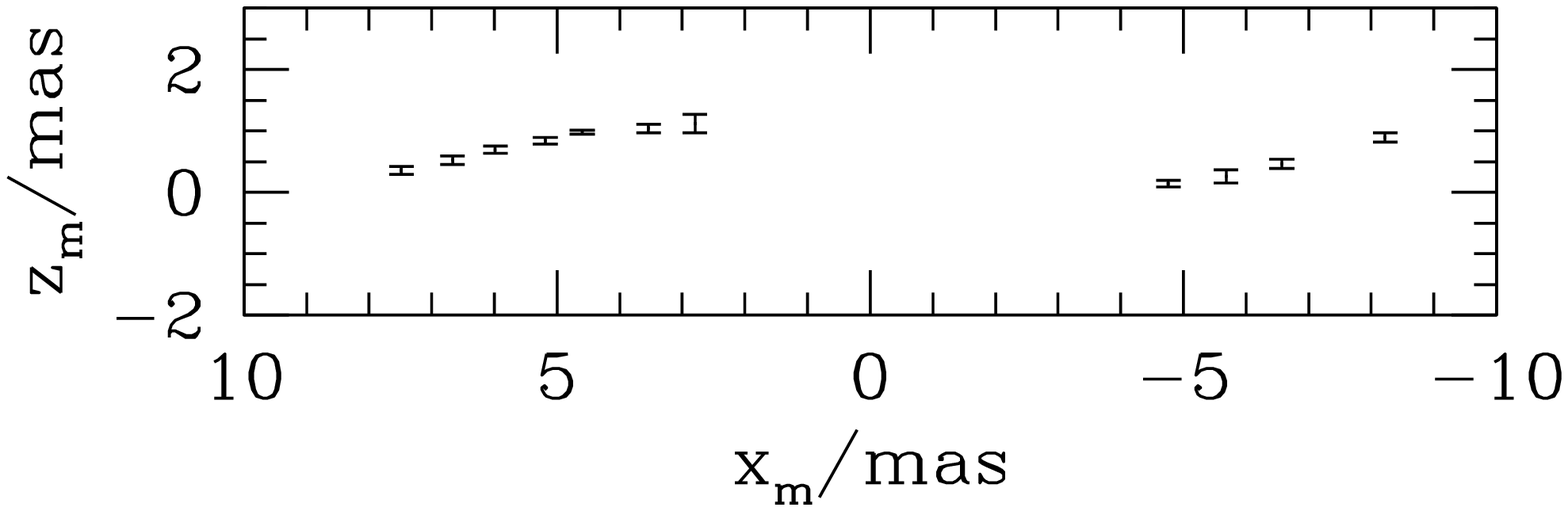}
\caption[]
{The maser distribution in space. The top plot shows all the data
  points and the bottom plot shows the binned data with error bars.}
\label{dataplot}
\end{figure}

In order to get an estimate of the measurement error we bin the maser
data points. We put the data into 11 bins and find the average
$\bar{x_i}$ and $\bar{z_i}$ in each where $i=1,2...11$. We then find
the standard deviations of each average point, $\sigma_{ix}$ and
$\sigma_{iz}$. We list these in Table~\ref{table2}.  In
Figure~\ref{dataplot} we plot all of the maser data in the top plot
and below we plot the binned data points with error bars.

The symmetry of the velocity distribution of the masers tells us that
the black hole is on the line $x_{\rm m}=0$ and we find the position
of the black hole, $z_{\rm add}$, from the symmetry of the shape of
the maser distribution.  We use the inner three points on the right in
the binned data and find an interpolated corresponding $z_{\rm m}$
value on the left hand side at the positive $x_{\rm m}$.  We can then
find the midpoint of these two $z_{\rm m}$ values from the left and
right hand sides. We average over the three points calculated and find
that $z_{\rm add}=0.52$ in the frame of the masers.

\begin{table*}
\begin{center}
\begin{tabular}{|l|l|l|l|l|l|}
\hline
$i$&$\bar{x_i}/\rm mas$&  $\bar{z_i}/\rm mas$& Number of Masers & $\sigma_{ix}$ & $\sigma_{iz}$\\ \hline
  1& -8.222 &  0.896 &  92 &  $8.018\times 10^{-2}$ &  $7.116\times 10^{-2}$  \\
   2& -6.574 &  0.465 &  90 &  $9.047\times 10^{-2}$ &  $7.836\times 10^{-2}$  \\
   3& -5.684 &  0.261 &  71 &  $5.939\times 10^{-2}$ &  $1.094\times 10^{-1}$  \\
   4& -4.760 &  0.143 &  24 &  $8.717\times 10^{-2}$ &  $5.445\times 10^{-2}$  \\
   5&  2.796 &  1.120 &  11 &  $1.271\times 10^{-1}$ &  $1.466\times 10^{-1}$  \\
   6&  3.539 &  1.041 &  19 &  $4.210\times 10^{-2}$ &  $7.498\times 10^{-2}$  \\
   7&  4.598 &  0.985 & 383 &  $7.121\times 10^{-2}$ &  $3.427\times 10^{-2}$  \\
   8&  5.188 &  0.843 &1042 &  $7.872\times 10^{-2}$ &  $5.321\times 10^{-2}$  \\
   9&  5.984 &  0.694 &1499 &  $1.510\times 10^{-1}$ &  $5.959\times 10^{-2}$  \\
   10&  6.666 &  0.528 & 358 &  $1.259\times 10^{-1}$ &  $6.753\times 10^{-2}$  \\
   11&  7.483 &  0.355 & 666 &  $1.484\times 10^{-1}$ &  $6.676\times 10^{-2}$  \\
  \hline
 \end{tabular}
\end{center}
\caption{The binned maser points with the number 
of points in each bin and the standard deviations.}
\label{table2}
\end{table*}

\subsection{Disc Section Transformation}

We have the shape of the disc section as given in the coordinates
$z_{\rm bh}$ and $x_{\rm bh}$. However, we can vary the size, the
rotation angle and position within the plane to get a good fit.  We
first let $R_0=R_{\rm warp}=1\,\rm mas$ in equation~(\ref{Wwarp}) and
find the disc section we want.

We rotate this disc section by an angle $q$ about the line of sight.
The black hole spin was originally in the north-south direction but
with this transformation it is at a position angle of $-q$. If we
assume that the observed jets are parallel to the spin of the black
hole, then the black hole spin should be at the same position angle as
the jets.

We can choose the factor $f=R_{\rm warp}/\rm mas$ which determines
where $R_{\rm warp}$ is relative to the maser distribution.  Our model
becomes
\begin{equation}
z_{\rm t}= \pm(z_{\rm bh} \cos q+x_{\rm bh}\sin q) f + z_{\rm add}
\end{equation}
and
\begin{equation}
x_{\rm t}= \pm(x_{\rm bh}\cos q-z_{\rm bh}\sin q) f ,
\end{equation}
where the sign depends on which side of the disc we are modelling.

We now have a model which can be rotated, stretched and moved to fit
the masers. We choose the parameters $\eta$, $f$ and $\beta$ for the
disc we wish to fit. Then we choose a disc section by choosing
$\phi_{\rm c}$ and $i_{\rm inc}$. We then rotate the disc section to
the best fit by varying $q$.

\subsection{Comparison of our Model to the Maser Data}

We can compare our analytical model to the binned points by finding
\begin{equation}
\frac{\chi^2_\nu}{\nu}=\frac{1}{\nu}\sum_{i=1}^N \frac{(z_i-z_{\rm t})^2}{\sigma_i^2},
\end{equation}
where $\nu=N-m$, $N=11$ is the number of bins and $m$ is the number of
parameters to be determined from the data.  We can change the
inclination of the black hole relative to the outer disc, $\eta$, and
also the power $\beta$.  We can rotate the disc about the axis of the
black hole by changing angle $\phi_{\rm c}$. We can also vary the
inclination, $i_{\rm inc}$, of the black hole to the line of sight. We can vary
the warp radius through $f$ .

For a given set of parameters for $\eta$, $\beta$, $f$, $i_{\rm inc}$,
$\phi_{\rm c}$, we minimise $\chi^2_{10}/10$ by varying $q$. We vary one
parameter at a time and so $m=1$ and $\nu=10$. The critical
probability values, $p$, for deviations of $1$, $2$ and $3\,\sigma$
for $\chi^2_{10}/10$ are given in Table~\ref{table}.  The probability of
exceeding the critical value of $\chi^2_{10}/10$ is $1-p$.

\begin{table}
\begin{center}
\begin{tabular}{|l|l|l|}
\hline
$\sigma$ & $p$   &  $\chi^2_{10}/10$   \\ \hline
1&  0.683   &  1.154 \\
2&  0.955   &  1.865 \\
3&  0.997   &  2.661 \\ \hline

 \end{tabular}
\end{center}
\caption{The probability of exceeding the critical value of $\chi^2_{10}/10$ is
$1-p$. We tabulate critical values of $\chi^2_{10}/10$ for deviations of $1$, $2$
and $3\,\sigma$.}
\label{table}
\end{table}

\begin{figure*}
  \centerline{\hbox{ \epsfxsize=8cm \epsffile{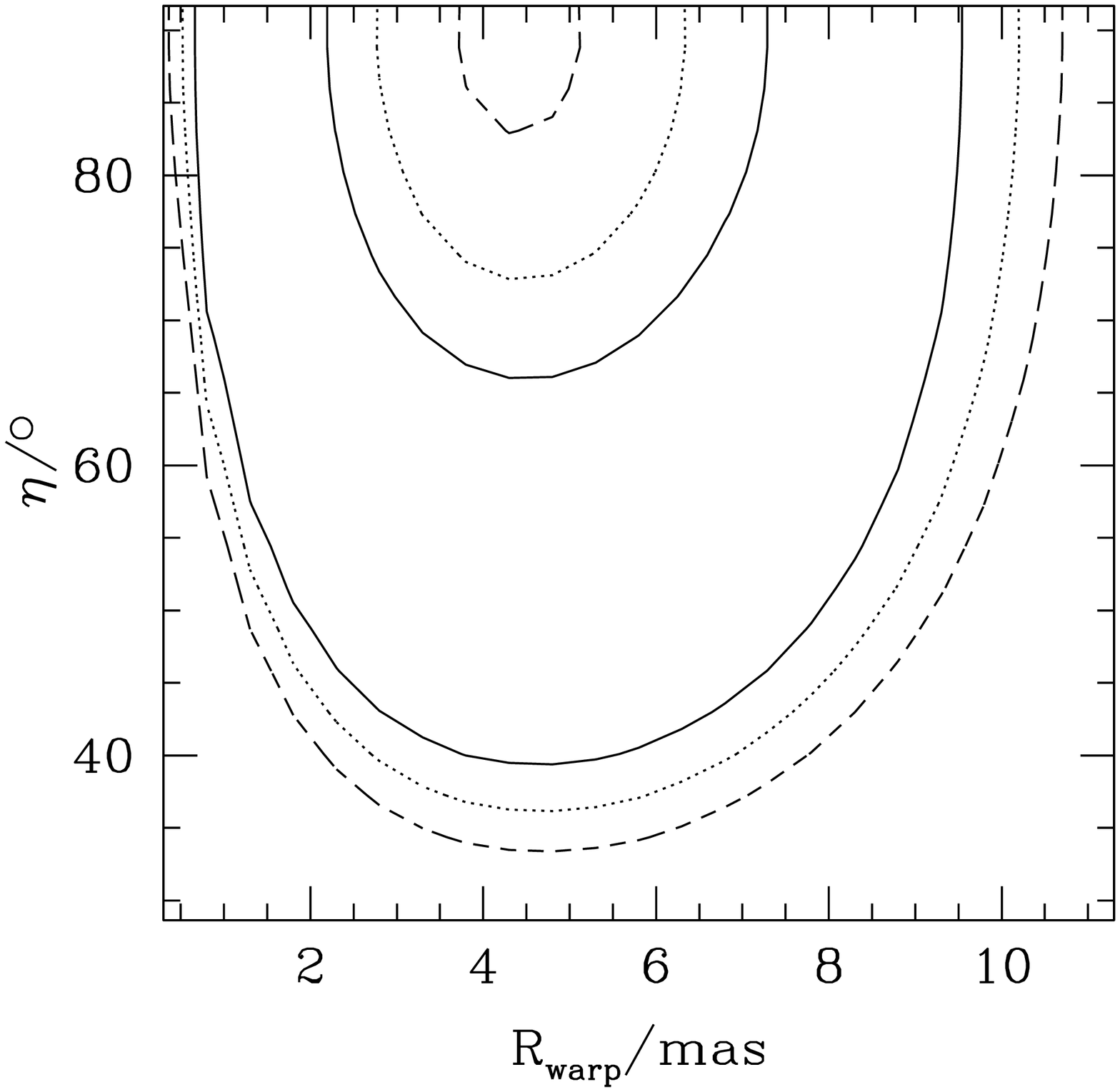}
      \epsfxsize=8cm \epsffile{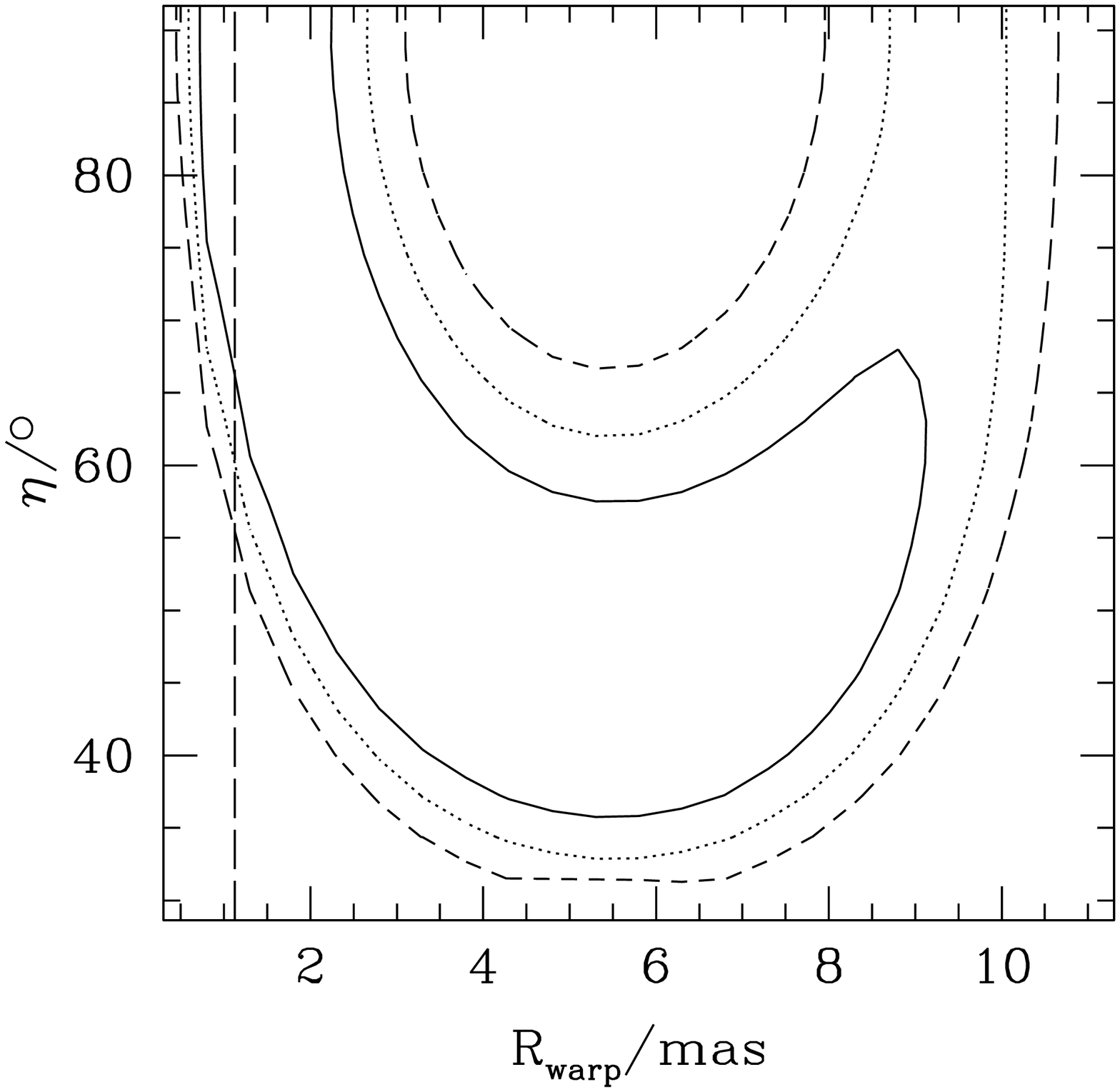} } }
  \centerline{\hbox{ \epsfxsize=8cm \epsffile{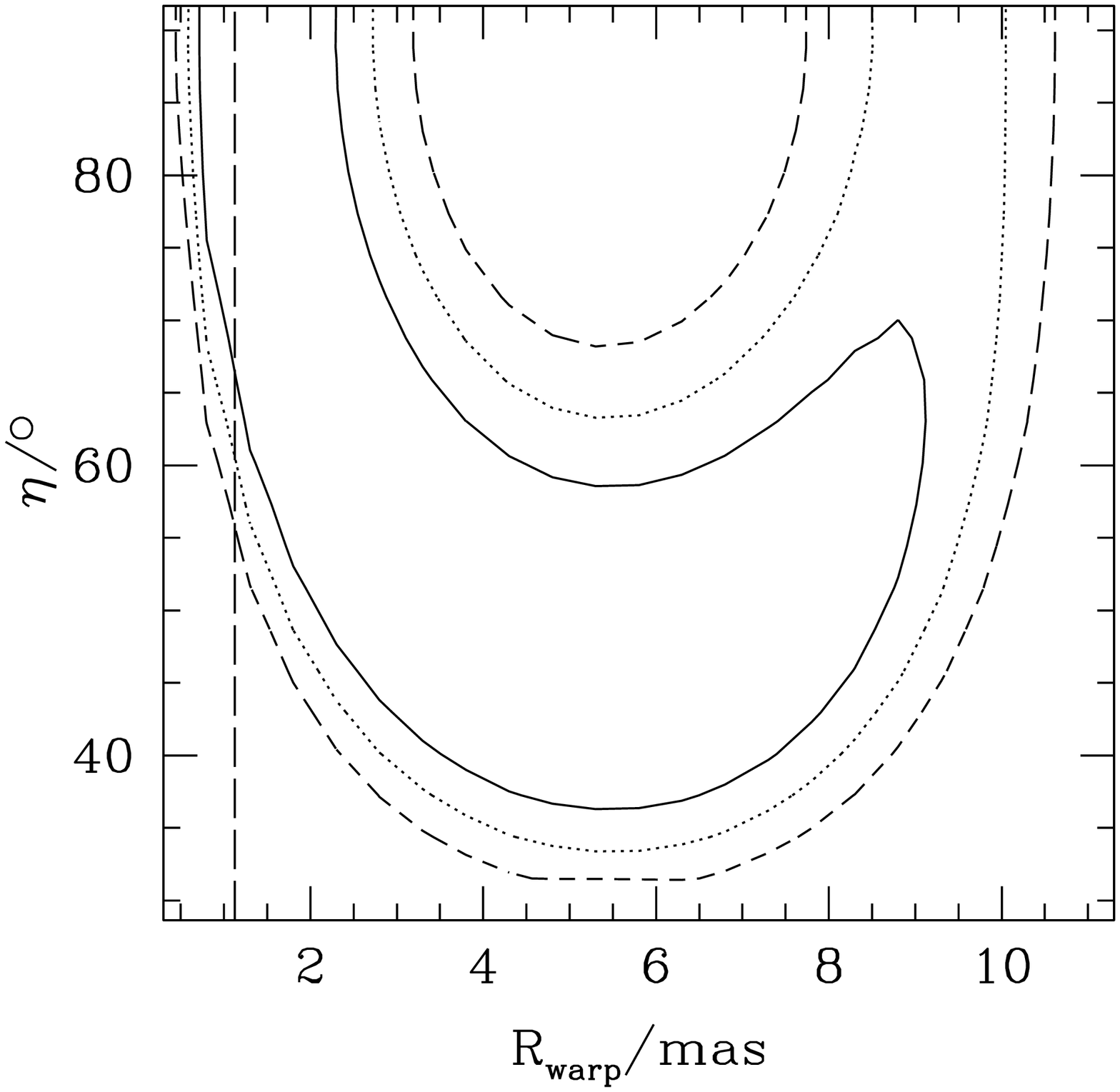}
      \epsfxsize=8cm \epsffile{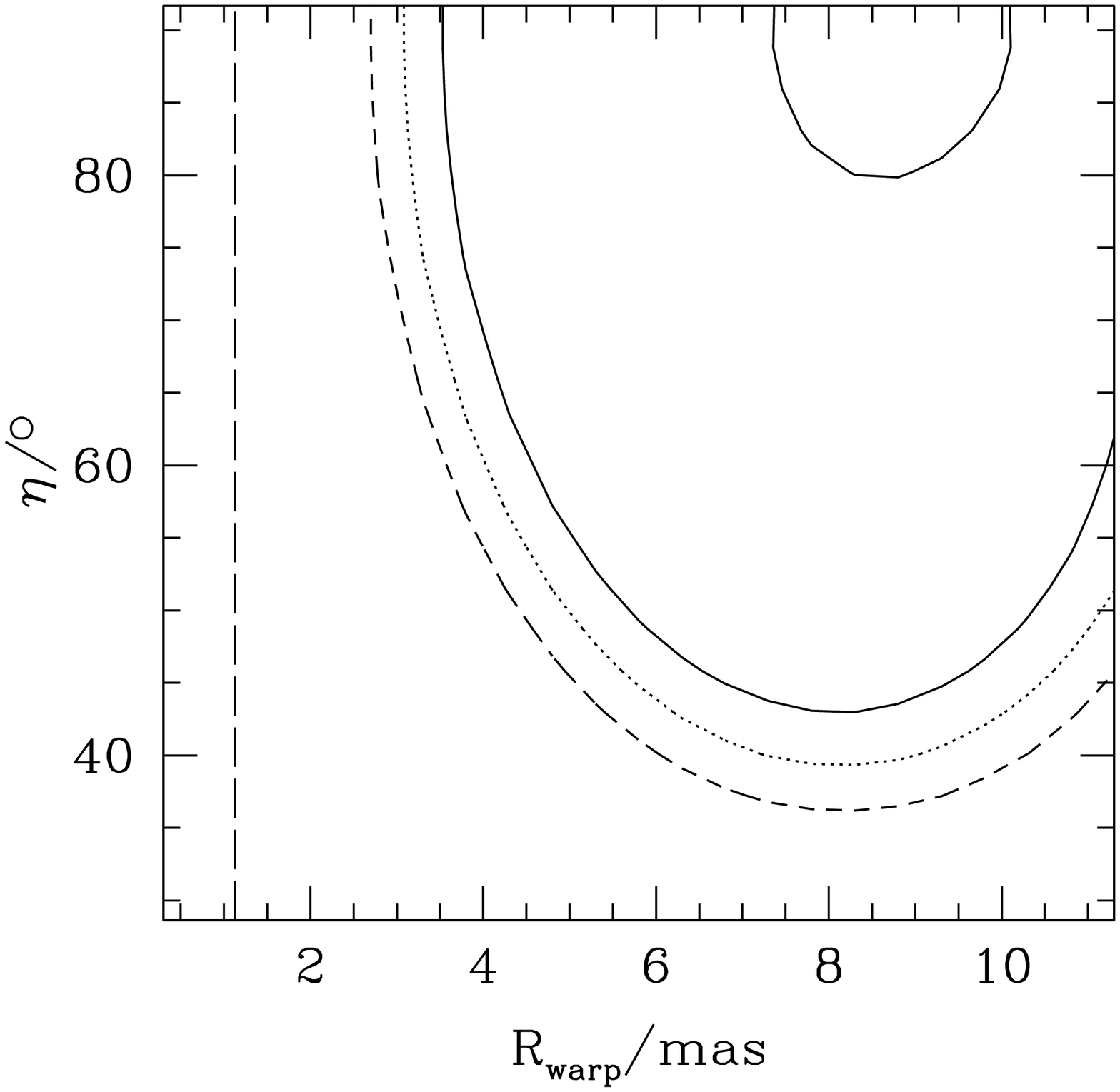}}}
\caption[]
{Contour plots of the values of $\chi^2_{10}/10$ for combinations of
  $R_{\rm warp}/\rm mas$ and $\eta$, the angle between the black hole
  and outer disc angular momenta.  The solid lines are
  $\chi^2_{10}/10=1.154$, the dotted lines are $\chi^2_{10}/10=1.865$
  and the dashed lines are $\chi^2_{10}/10=2.661$ which correspond to
  $1$, $2$ and $3\, \sigma$ deviations.  The top left plot has
  $\beta=1$, $\phi_{\rm c}=0$ and $i_{\rm inc}=0$.  The top right plot
  has $\beta=2$, $\phi_{\rm c}=0$ and $i_{\rm inc}=0$.  The bottom
  left plot has $\beta=2$, $\phi_{\rm c}=0$ and $i_{\rm
    inc}=10^\circ$.  The bottom right plot has $\beta=2$, $\phi_{\rm
    c}=45^\circ$ and $i_{\rm inc}=0$. The vertical long dashed lines
  shown for the cases with $\beta=2$ are the upper limit on $R_{\rm
    warp}$ derived from the condition that the disc must be stable
  against self gravity.}
\label{contours}
\end{figure*}

In Figure~\ref{contours} we plot contours of $\chi^2_{10}/10$ in $f$ and
$\eta$ space for different values of $\beta$, $\phi_{\rm c}$ and $i$.
Each point in the plot is the best fitting point when we vary $q$. The
plots would be symmetric over the line $\eta=90^\circ$ if we were to
extend them to higher $\eta$ because the disc shape is the same for
$\eta$ and $\pi-\eta$ but if $\eta>\pi/2$ the disc is counter aligned.
We see that we can fit the disc shape for a variety of parameters.

Generally, we see that we need $\eta>30^\circ$ or so for a reasonable
fit and the warp radius must be on the scale of the masers.  Comparing
the top plots which have different $\beta$ only, we see that for the
lower value of $\beta=1$ we can fit higher $R_{\rm warp}$ better but
we cannot fit such small $\eta$ as for $\beta=2$.

The bottom left plot is the same as the top right plot except the
black hole is inclined at an angle of $i_{\rm inc}=10^\circ$ to the plane of the
sky.  Comparing this to the top right plot which has $i_{\rm inc}=0$ we see that
the the range of parameters we can fit is not changed much for small
inclinations of the black hole from the plane of the sky. Note that
the method we have used to generate these fits does not take into
account whether masers would actually be seen for the inclinations
chosen. We need the inclination of the black hole from the plane of
the sky to be small in order to see masers at all.

In all the plots considered so far, we have chosen the disc section of
a given disc with the most warping, the one where the plane of the sky
contains the angular momentum of the black hole and outer disc
vectors. In the bottom right plot we choose a different disc section
with $\phi_{\rm c}=45^\circ$. Here, we cannot fit such small $R_{\rm
  warp}$ or $\eta$.
 
In Section~\ref{sec:rwarp} we found an upper limit on the warp radius
for the disc to be stable against self gravity that $R_{\rm warp}$
must be less that $1.65\,\rm mas$. This rules out the case with
$\phi_{\rm c}=45^\circ$ completely. We find that we can only fit discs
which have $|\phi_{\rm c}|<25^\circ$. In order for our models to
explain the warping in NGC 4258, the black hole and outer disc angular
momentum vectors must be close to lying in the the plane of the sky.
By the bottom left plot, the limit also implies that we need
$\eta>60^\circ$.

\subsection{Best Fitting Models}

\begin{figure}
\epsfxsize=8.4cm 
\epsfbox{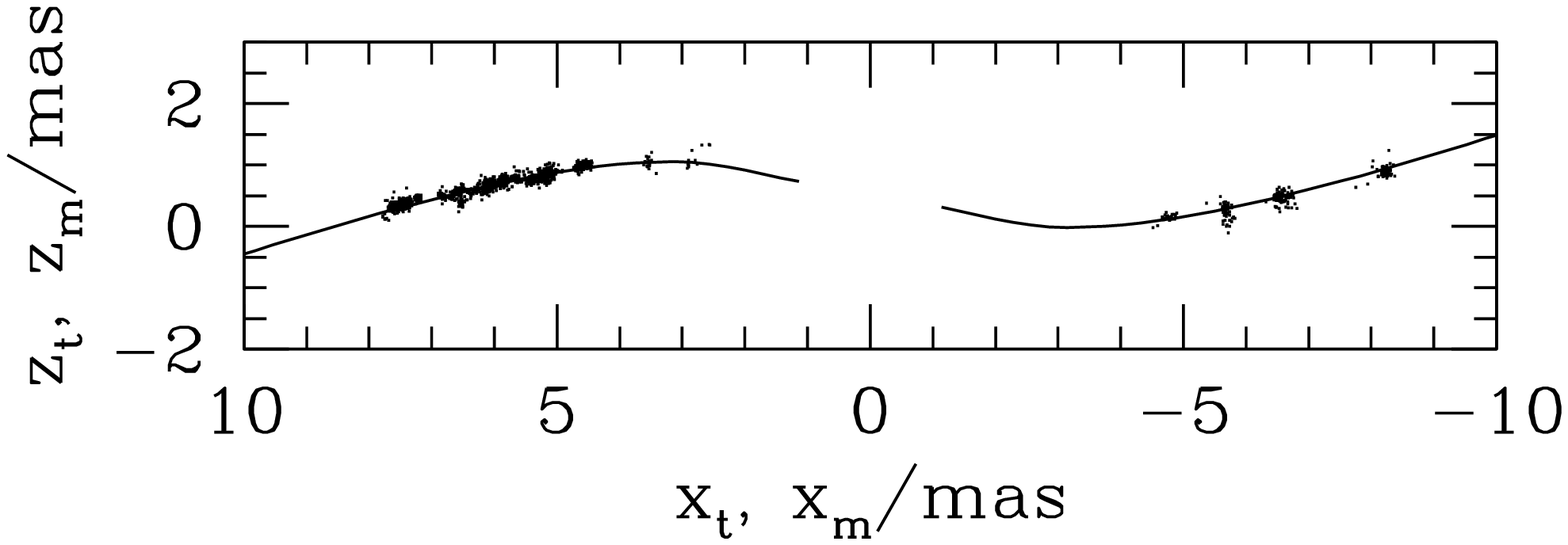}
\epsfxsize=8.4cm 
\epsfbox{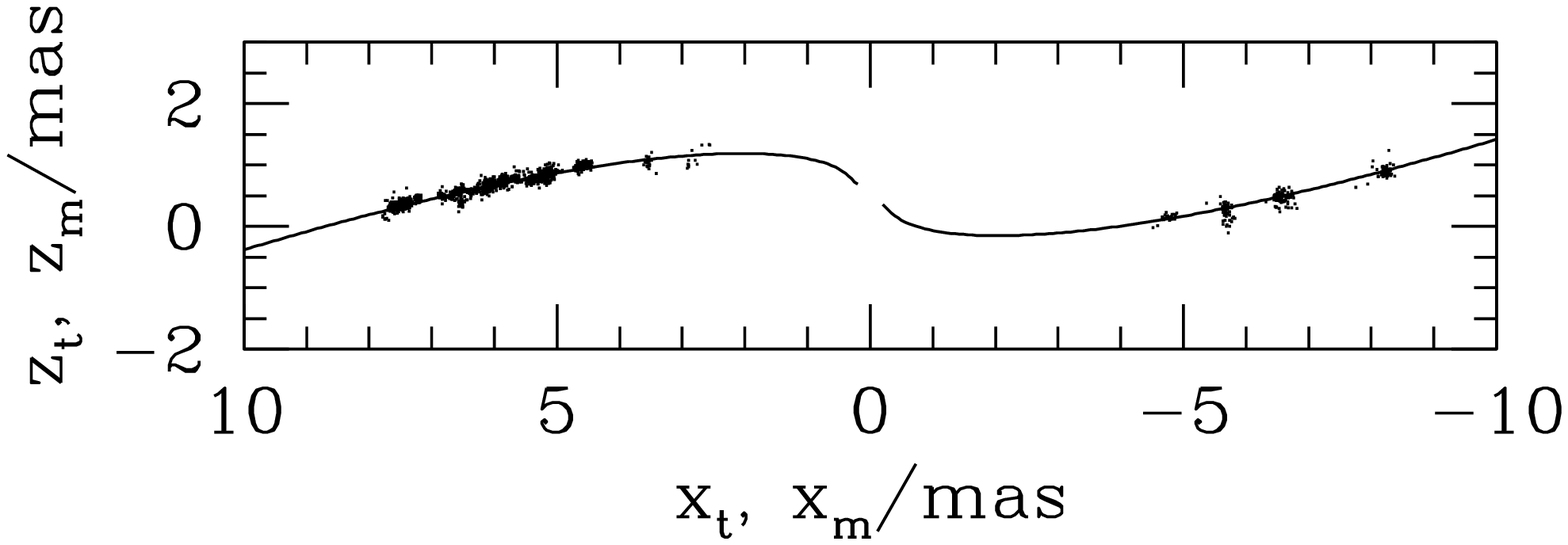}
\caption[]
{The maser points with the best fitting models with $\beta=2$, $i_{\rm inc}=0$
  and $\phi_{\rm c}=0$. The top plot has $\chi^2_{10}/10=0.173$,
  $\eta=45.8^\circ$, $R_{\rm warp}=5.8\,\rm mas$ and $q=10.6^\circ$.
  The bottom plot is the best fitting model which is stable against
  self gravity and has $\chi^2_{10}/10=0.203$, $\eta=90^\circ$, $R_{\rm
    warp}=1.3\,\rm mas$ and $q=39.8^\circ$. }
\label{model}
\end{figure}

We consider the best fitting combination of parameters for the case
$\beta=2$, $i=0$ and $\phi_{\rm c}=0$. The parameters giving the
smallest $\chi^2_{10}/10$ of $0.173$ are $\eta=45.8^\circ$ and $R_{\rm
  warp}=5.8\,\rm mas$ and this has $q=10.6^\circ$. The position angle
of the black hole is $-q$ and so if we assume that the jets are
parallel to the spin of the black hole then the position angle of the
jets here is $-10.6^\circ$. This is on the edge of our error for the
observed position angle of the jets. The inclination of the disc at
$R=3.9\,\rm mas$ to the plane of the sky is $\zeta_{\rm
  in}=7.4^\circ$.  In the top of Figure~\ref{model} we plot this fit
with the maser data points. For this warp radius we would need
$H/R=0.0002$ and so the disc would be unstable against self gravity.

The best fitting model which is stable against self gravity has
$R_{\rm warp}=1.3\,\rm mas$, $\eta=85.9^\circ$,
$\chi^2_{10}/10=0.20331$ and $q=39.6^\circ$ and we plot this in the
bottom plot in Figure~\ref{model}. If we take into account our
constraint on $R_{\rm warp}$ then we need a larger angle $q$ which
means that the black hole has position angle which is more negative
and not within the error on the observed position angle of the jets.
However, the inclination of the disc at $R=7\,\rm mas$ to the plane
of the sky is $\zeta_{\rm in}=8.6^\circ$ which is closer to the
observed value of $8.4^\circ$.

The angles of the jets are very poorly determined and vary in time
indicating that the angle is affected by the surrounding medium. None
of the models that fit the data well have negative $q$.  Once we
constrain the warp radius we find we need an even larger $q$ to fit
the data.  However, the jet is not very clearly observed \citep{H97}
and so we cannot rule out our models because of this.

\section{Physical Implications for the Disc}
\label{timescales}

We now consider the physical implications of our fit to the disc
structure in NGC 4258.  First we calculate the viscous timescale in
the disc.  For $R_{\rm warp}=0.83\,\rm mas$ with $H/R=0.001$ we find
the viscosity to be
\begin{align}
  \nu_1 & = 4.3\times 10^{18}
  \left(\frac{R}{R_{\rm warp}}\right)^{\beta}\,\rm cm^2 s^{-1}.
\end{align}
The viscous timescale in the disc for $\beta=2$ is independent of $R$
and is given by
\begin{align}
t_{\nu_1}= & \frac{R^2}{\nu_1}=  \frac{R_{\rm warp}^2}{\nu_{10}}= 6.3 \times 10^7\,\rm yr.
\end{align}
and represents the timescale of the movement of mass through the disc.
The second viscosity timescale is
\begin{equation}
t_{\nu_2}= \frac{\alpha_{10}}{\alpha_{20}} t_{\nu_1} =6.3\times 10^6\,\rm yr
\end{equation}
and this is the timescale associated with the generation of the warp
in the disc.

The precession timescale due to the Lense-Thirring effect is given by
\begin{equation}
t_{\rm LT}=\frac{c^2R^3}{2GJ}=\frac{c^3R^3}{2aG^2M_{\rm BH}^2}.
\end{equation}
At $R_{\rm warp}$ we have $t_{\rm LT}=t_{\nu_2}$ because of the
definition of $R_{\rm warp}$.  At $R_{\rm in}$, $t_{\rm LT}=4.6\times
10^8\,\rm yr$ which is longer than $t_{\nu_1}$. Hence, for the disc to
be in a steady state we require it to last for much longer than the
viscous timescale. This means that the disc must be effectively of
infinite mass because it is continually supplied with matter.

Seyfert outbursts are stochastic randomly-orientated, short-lived
accretion events. If this system is a Seyfert system then the maser
disc is just the outer regions of a disc generated by an accretion
event.  Seyfert episodes thought to be shorter than $10^6\,\rm yr$
\citep{S84} and nuclei evolve through at least 100 of these. The disc
becomes stable and thin in a few orbital periods, a time of around
$10^3\,\rm yr$ at the outer edge of the disc. The warp then just
represents the distribution of angular momentum in the accreted
matter. The time to change the warp, around $10^8\,\rm yr$, is much
longer than the length of the Seyfert outburst and the typical time to
the next outburst. It is hard to reconcile our model without upsetting
this understanding of Seyfert outbursts.

However NGC4258 is a rather inactive AGN. The assumed luminosity is
well below the Eddington luminosity. We suggest that it could be in a
more quiescent phase with steady accretion.

\section{Conclusions}

We can fit the shape of the warped disc in NGC 4258 with steady state
accretion disc which has a misaligned spinning black hole that causes
the disc to warp by Lense-Thirring precession. We need the angle
between the black hole and the outer disc to be greater than
$60^\circ$ in order to get a good fit.  We need the warp radius to be
$1<R_{\rm warp}/{\rm mas}<10$. We need the angular momentum of the
outer disc to be at an angles of less than $25^\circ$ to the plane of
the sky. The black hole spin must be close to the plane of the sky to
see masers.

The best fitting model has the warp radius at $5.8\,\rm mas$ and a
black hole position angle of $-10^\circ$ but this disc would be
unstable to self gravity. The best fitting model which would not
be self gravitating has $R_{\rm warp}=1.3\,\rm mas$ but the position
angle of the black hole is $-40^\circ$.  The observed direction of the
jet is $5^\circ\pm 15^\circ$ and so cannot be easily accounted for
using this model.

We need the disc to be very long lived in order for it to become
warped in its lifetime.  If the disc was in steady state all the way
in, it would be unstable against self-gravity.  The requirements of
the model are hard to fit with our understanding of Seyfert outbursts
which are thought to last only $10^6\,\rm yr$.  However, it remains
that we can fit the observed maser distribution with an accretion disc
warped by Lense-Thirring precession remarkably well.

\section*{Acknowledgements}

I thank Jim Pringle for many useful discussions and reading the
manuscript and Christopher Tout for help with statistics and other
useful discussions. I thank Alice Argon and Jim Moran for advice about
the maser observations and Anderson Caproni and Lincoln Greenhill for
useful comments.

\label{lastpage}
\end{document}